\documentclass[galaxies,article,submit,moreauthors,pdftex, biblatex, 10pt,a4paper]{mdpi}

\firstpage{1} 
\articlenumber{x}
\doinum{10.3390/------}
\pubvolume{xx}
\pubyear{2016}
\copyrightyear{2016}
\externaleditor{Academic Editor: }
\history{Received: date; Accepted: date; Published: date}
 \theoremstyle{mdpi}
 \newcounter{thm}
 \newcounter{ex}
 \newcounter{re}
 \theoremstyle{mdpidefinition}

\Title{Multi-wavelength Intra-day Variability and Quasi-periodic Oscillation in Blazars}
\Author{Alok C.\ Gupta}
\AuthorNames{Firstname Lastname, Firstname Lastname and Firstname Lastname}

\address{%
Aryabhatta Research Institute of Observational Sciences (ARIES), Manora Peak, Nainital -- 263002, India}
\corres{Correspondence: acgupta30@gmail.com}

\abstract{We reviewed multi-wavelength blazars variability and detection of quasi-periodic oscillations
on intra-day timescales. The variability timescale from few minutes to up to less than a days is commonly known 
as intra-day variability. These fast variations are extremely useful to constrain the size of emitting region,
black hole mass estimation, etc. It is noticed that in general blazars show intra-day variability in the
complete electromagnetic spectrum. But some class of blazars either do not show or show very little intra-day 
variability in a specific band of electromagnetic spectrum. Blazars show rarely quasi-periodic oscillations
in time series data in optical and X-ray bands. Other properties and emission mechanism of blazars are also
briefly discussed.}      


\begin{document}

\section{Introduction}

\noindent
It is commonly accepted that super massive black holes (SMBHs, with masses between 
10$^{6} -$ 10$^{10}$ M$_{\odot}$) are present in the nuclei of all galaxies with stellar bulges. At any 
given time a few percent of these SMBHs are fed with a sufficient amount of gas that they will possess 
significant accretion discs. The emission of these discs is comparable to the  total emission of  stars 
in the entire host galaxy because of a very high efficiency for the conversion of matter into radiation 
as it spirals into a BH. This is the fundamental mechanism underlying ``active galactic nucleus'', or AGN. \\
\\
Roughly 85$-$90 \% of AGNs have very little radio emission ($F_{5GHz} / F_{B} \leq$ 10, here F$_{5GHz} =$ flux
at radio 5 GHz and F$_{B} =$ flux at optical $B$ band 4400\AA) and are therefore called radio-quiet AGNs (RQAGNs). 
The remaining $\sim$ 10--15 \% of AGNs are radio-loud AGNs (RLAGNs). It has been proposed that different types of 
AGNs can be explained by the idea that different line of sight (LOS) angles can play an important role in 
understanding their different properties (Antonucci 1993; Urry \& Padovani 1995). \\ 
\\
Blazars belongs to the RLAGN class and its LOS is pointed close to the observer. Blazars show rapid variability 
at almost all wavelengths of the electromagnetic (EM) spectrum with the emission being strongly polarized 
(optical linear polarization $\geq$ 3\%). Due to their strong and large amplitude variability nature in the 
complete EM spectrum, they are considered as transient astronomical objects. BL Lacertae objects (BLLs) and 
flat spectrum radio quasars (FSRQs) are collectively known as blazars. BLLs show featureless optical continua 
(no prominent emission or absorption lines) while FSRQs show prominent emission lines in their optical spectra.
The radiation from blazars is dominated by non-thermal emission at all wavelengths, consisting of two broad
spectral bumps (Ghisellini et al. 1997; Fossati et al. 1998): A low-frequency component from radio to the UV 
or X-rays, generally agreed to be due to synchrotron radiation from relativistic electrons in the jet, and a 
high-frequency component from X-rays to $\gamma-$rays, which can be either due to Compton scattering of 
lower-frequency radiation by the same relativistic electrons (leptonic models i.e. Krawczynski 2004) or due 
to interactions of ultra-relativistic protons in the jet (hadronic models), either via proton synchrotron 
radiation (Mucke et al. 2003) or via secondary emission from photo-pion and photo-pair production process 
(B$\ddot{o}$ttcher 2007; and references therein). \\
\\
Blazars can be classified into three sub-classes, depending on the peak frequency of their synchrotron emission:
LSPs (low-synchrotron-peaked blazars), consisting predominantly of LBLs (red or low energy or radio selected
blazars) and defined by a peak of their synchrotron component at $\nu_{sy} <$ 10$^{14}$ Hz, ISPs
(intermediate-synchrotron-peaked blazars) or IBLs, consisting mostly of intermediate blazars, defined by
10$^{14}$ Hz $< \nu_{sy} <$ 10$^{15}$ Hz, and HSPs (high-synchrotron-peaked blazars), all of which are HBLs
(blue or high energy or X-ray selected blazars) and which are defined through $\nu_{sy} >$ 10$^{15}$~Hz 
(Abdo et al. 2010).
The high-energy component of the spectral energy distribution (SED) of blazars extends up to $\gamma-$rays,
peaking at GeV energies in LSPs and at TeV energies in HSPs. Blazar properties are consistent with relativistic
beaming, i.e. bulk relativistic motion of the jet plasma at small angles to the line of sight, which gives
rise to a strong amplification and rapid variability in the observer's frame. \\
\\
The study of variability is one of the most powerful tools for understanding the nature and processes
occurring in blazars. Variability of blazars can be broadly divided into three classes. Significant variations
in flux may occur over few tens of minutes to the course of less than a day, often called micro-variability,
intra-night variability or intra-day variability (IDV) (Wagner \& Witzel 1995). Short term variability (STV) 
can range time scales from days to few months and long term variability (LTV) can have time scales of several 
months to several years (Gupta et al. 2004). \\
\\
Variability observations on IDV timescales in AGN is the most puzzling. Variability of blazars on IDV 
time scale can provide important clues to the physics of the
innermost nuclear regions in these objects. Blazar properties are consistent with relativistic beaming
caused by bulk relativistic motion of the jet plasma at small angles to the line of sight, which gives
rise to a strong amplification and rapid variability in the observer's frame. 
With simultaneous multi-wavelength observations of blazars in the entire EM spectrum is an important 
tool to test several The possible models for IDV: shock-in-jet models, accretion-disk based models, and 
models based on plasma instabilities in shear layers, etc. 

\section{Intra-day Variability in different EM Bands}

\noindent
There have been several dedicated monitoring campaigns in which the IDV of blazars has been studied over 
the entire EM spectrum (e.g. Miller et al. 1989; Carini et al. 1990, 1991, 1992; Carini \& Miller 1992; 
Quirrenbach, et al. 1991;  Heidt \& Wagner 1996, 1998; Sagar et al. 1999; Montagni et al. 2006; Aharonian 
et al. 2007; Gupta et al. 2008a, 2008b, 2012, 2016, 2017; Gaur et al. 2010, 2012a, 2012b, 2015; Foschini 
et al. 2011; Kushwaha et al. 2014; Paliya et al. 2015; Paliya 2015; Kalita et al. 2015; Agarwal \& Gupta 
2015; Agarwal et al. 2015, 2016; Ackermann et al. 2016;  Pandey et al. 2017; Aggrawal et al. 2017; and 
references therein). \\
\\
There are several methods which can be used to find the genuine IDV and variability parameters in time 
series data. These methods are described by different groups and relevant source references are cited in 
their papers. Romero et al. (1999) introduced C--Test which is used by several groups (e.g. Jang \& Miller 
1997; Stalin et al. 2004; Gupta et al. 2008a; Gaur et al. 2012b; Agarwal \& Gupta 2015; and references 
therein). But later de Diego (2010) explained that the C--Test statistic is too conservative approach for 
IDV detection. IDV detection results can be tested using F--test which is a distributed statistic (de 
Diego 2010). F--test is also used by several groups (e.g. Gaur et al. 2012b; Agarwal \& Gupta 2015; and 
references therein). A better method called power enhanced F--test is introduced by (de Diego 2014; de Diego
et al. 2015). This method takes care of large brightness difference in blazar and comparison / standard stars 
or large brightness difference in comparison / standard stars. It is used by Gaur et al. (2015). $\chi^{2}-$Test
is explained by (Gaur et al. 2012b; Agarwal \& Gupta 2015) also provide the presence of IDV in time series data.
The analysis of variance (ANOVA) test is a very robust test to investigate the variability in the light 
curves (LCs) 
(e.g. de Diego et al. 1998; Gaur et al. 2012b; Agarwal \& Gupta 2015; Gupta et al. 2017; and references therein). 
Percentage amplitude variation in the time series data is introduced by Heidt \& Wagner (1996) and used in several 
papers (e.g. Gupta et al. 2008a; Gaur et al. 2012b; Agarwal \& Gupta 2015; Gupta et al. 2017; and references 
therein). One another method called excess variance and fractional rms variability amplitude are introduced by 
(Edelson et al. 2002; Vaughan et al. 2003) and extensively used for getting the variability in AGN LCs 
(e.g. Kalita et al. 2015; Bhagwan et al. 2016; Gupta et al. 2016; Pandey et al. 2017; and references therein). \\
\\
IDV in blazars can be intrinsic to the source or due to extrinsic origin. Interstellar scintillation and 
gravitational microlensing are the main extrinsic cause of IDV. Interstellar scintillation is only relevant
in low-frequency radio observations. Gravitational microlensing is only applicable in a few blazars which 
are lensed system e.g. the blazar AO 0235+135 at z = 0.94  have revealed foreground-absorbing systems at 
z = 0.524 and z = 0.851 (Cohen et al. 1987; Nilsson et al. 1996). In the blazars where IDV is detected in 
low-frequency radio observations or gravitationally lensed sources can also have some intrinsic origin, and
to find that we need to do simultaneous multi-wavelength of the blazar, and co-related variability in 
different EM bands will be helpful to find the nature of variability. \\
\\
IDV in the high state (pre/post to outburst state) of blazars can be explained by jet based models e.g. helical 
instabilities in the jet (Marscher \& Gear 1985; Qian et al. 1991) or turbulence behind the shock in the jet 
(Marscher, Gear, Travis 1992). Jet based models can explain IDV over the entire range of EM wavelengths. Other 
theoretical models seek to explain IDV in blazars (mainly in their low-state) involving accretion-disk based 
models. These models include pulsations of the gravitational modes of the gaseous disk (Kato \& Fukue 1980; 
Nowak \& Wagoner 1992) or orbital signatures from ``hot-spots'' in the gas surrounding the black hole, either 
from the disk itself or the corona above it (Zhang \& Bao 1991; Mangalam \& Wiita 1993). Accretion-disk based 
models can explain the variations in optical, UV and X-ray bands, but are difficult to connect to the observed 
rapid variability in $\gamma-$rays. Plasma instabilities in the jet (e.g., Kelvin-Helmholtz type instabilities 
due to the interaction of a fast inner spine of the jet with a slower, outer layer) could play an important role 
in the production of IDV at a variety of wavelengths.

\subsection{IDV In Gamma-rays}

\noindent
Most of the ground and space based gamma-ray experiments are not sensitive enough to observe blazars
with few minutes time resolution. So, it is extremely difficult to do IDV studies of blazars. But there
is an excellent IDV observation of the blazar PKS 2155--304 in an exceptional very high energy gamma-ray
flare observed on July 28, 2006 using HESS (High Energy Stereoscopic System). The IDV LC obtained with 
time resolution of 1 minute and IDV is seen up to 10 minutes which gives Doppler factor more than 100. 
The average flux at $>$ 200 GeV during outburst was $\sim$ 7 times of flux observed from Crab Nebula 
(Aharonian et al. 2007). Thanks to the LAT (large area telescope) on board to Fermi gamma-ray space
telescope (hereafter Fermi-LAT; Atwood et al. 2009) which is observing sky in gamma-ray since its launch
in June 2008. Fermi-LAT covers 20 MeV -- 300 GeV energies and made a revolution in the discovery of blazars.
It has observed several blazars in flaring state and detected strong IDV (Foschini et al. 2011; Kushwaha 
et al. 2014; Paliya et al. 2015; Paliya 2015; Ackermann et al. 2016; and references therein). Fermi-LAT 
LC with time resolution of minutes of blazars were presented by (Foschini et al. 2011; Ackermann 
et al. 2016) while (Kushwaha et al. 2014; Paliya et al. 2015; Paliya 2015) presented Fermi-LAT LCs 
of blazars with few hours time bin. Kushwaha et al. (2014) reported Fermi-LAT LC of the blazar 
PKS 1222+216 with 6h time bin show asymmetric rise profiles but rapid decline during the April 2010 flare.      

\subsection{IDV in X-rays}

\noindent
A pilot project on searching for IDV in blazars are initiated (Gaur et al. 2010; Kalita et al. 2015; Gupta
et al. 2016; Pandey et al. 2017; Aggrawal et al. 2017; and references therein). A sample of four HBLs observed 
on 23 occasions by XMM--Newton are studied, IDV timescales ranging from 15.7 to 46.8 ks were found on 8 
occasions, in 13 cases IDV timescales were longer than the data length, and hint of weak quasi periodic 
oscillations (QPOs) were observed on one LC each of blazars ON 231 and PKS 2155 -- 304 (Gaur et al. 2010).
Gonz$\acute{a}$lez-Mart$\acute{i}$n \& Vaughan (2012) took the sample of 104 nearby AGN (z $<$ 0.4) observed 
with XMM-Newton. The AGN sample also include several blazars. They did power-spectrum density (PSD) analysis 
of all those LCs and reported IDV variability parameters. The LBL 3C 273 observed on 24 occasions during 
2000 -- 2012 with XMM-Newton, and on IDV time-scales 3C 273 have shown occasionally small amplitude variability 
(Kalita et al. 2015). A complete sample of 12 LBL and IBL with 50 IDV LCs taken with XMM-Newton are compiled 
(Kalita et al. 2015; Gupta et al. 2016b). It is noticed that
the duty cycle of genuine IDV detection in LBL and IBL in X-ray band is only 4\% (2 out of 50 LCs) (Gupta et al.
2016). It is concluded that probably peak of the spectral energy distribution seems to be responsible for IDV 
properties (Gupta et al. 2016b). In a recent study (Pandey et al. 2017) reported search for X-ray IDV in five TeV
blazars using NuStar. Four TeV blazars have shown large amplitude IDV, using auto correlation function (ACF),
IDV timescale in the range of 2.5 to 32.8 ks is reported in eight LCs of Mrk 421; a timescale of about 8.0 ks
for one LC in Mrk 501; and timescales of 29.6 to 57.4 ks in two LCs of PKS 2155--304. In general soft (3 -- 10 keV)
and hard (10 -- 79 keV) LCs were well correlated which indicate the same population is emitting soft and hard
X-rays. IDV timescales are used to calculated $\delta$ the Doppler factor, $B$ the magnetic field, $\gamma$
the Lorentz factor, and $R$ the size of emitting region (Pandey et al. 2017). Recently in another search for
X-ray IDV, 83 LCs of the TeV blazar Mrk 421 taken during 1999 -- 2015 with {\it Chandra} are studied (Aggrawal 
et al. 2017). IDV timescale ranging 2.4 to 30.0 ks, IDV duty cycle $\sim$ 77\%, soft (0.3 -- 2.0 keV) and hard 
(2.0 -- 10.0 keV) LCs were well correlated with zero lag are found. IDV timescales are also used to calculate 
$\delta$ the Doppler factor, $B$ the magnetic field, $\gamma$ the Lorentz factor, and $R$ the size of emitting 
region for the blazar Mrk 421 (Aggrawal et al. 2017).     

\subsection{IDV in Optical and infrared (IR) bands}

\noindent
The first evidence of optical micro-variability is reported in BL Lacertae by Miller et al. (1989). The 
result of this paper motivated several groups around the globe to start dedicated project to search for 
optical micro-variability in blazars. Optical IDV in blazars is pioneered by the USA group in which they 
studied optical micro-variability in five blazars (e.g. Miller et al. 1989; Carini et al. 1990, 1991, 
1992; Carini \& Miller 1992), and reported that the probability of finding genuine micro-variability is 
about 80\% for the blazar continuously monitored for $<$ 8 hours. An extensive search for optical IDV in 
a sample of 34 BLLs from 1 Jy catalog was done by (Heidt \& Wagner 1996). IDV was detected in 28 out of 
34 BLLs (82\%), and 75\% of the variable BLLs changed significantly over a time span $<$ 6 hours. But this 
data lacks continuity in the LCs. Gupta \& Joshi (2005) compiled the optical IDV studies of blazars till 
$\sim$ 2004 and noticed that the occurrence of IDV on blazars if observed for less than 6 h is about 60–65\%. 
If the blazar is observed for more than 6h then the possibility of IDV detection is about 80–85\%     
Several hundred nights of optical IDV search in blazars were done by different Chinese groups (e.g. Bai 
et al. 1998, 1999; Dai et al. 2001; Xie et al. 1999, 2001, 2002, 2004; 
Fan et al. 2001, 2004; Qian \& Tao 2004; Wu et al. 2005, 2007; Poon et al. 2009; Hu et al. 2014; Li et al. 
2017; Xiong et al. 2016; Guo et al. 2017; Feng et al. 2017; and references therein). Significant IDV is 
detected in most of these observations but in several papers data lacks continuity in the LCs. Indian 
groups also did extensive search for optical/IR IDV in blazars with collaborators around the globe (e.g.
Sagar et al. 1999; Ghosh et al. 2000, 2001; Gupta et al. 2004, 2008a, 2008b, 2016a, 2017; Stalin et al. 
2005, 2006; Goyal et al. 2009; Rani et al. 2011; Gaur et al. 2012a, 2012b, 2012c, 2015; Agarwal \& Gupta 
2015; Agarwal et al. 2015, 2016; and references therein). There are some other collaborations world wide 
which are doing search for optical/NIR IDV of blazars (e.g. Heidt \& Wagner 1998; Villata et al. 2000,
2002, 2004, 2008; Romero et al. 2002; Papadakis et al. 2003, 2004; Ostorero et al. 2006; Xilouris et al. 
2006; Bachev et al. 2011, 2012, 2017; Bachev 2015; Bhatta et al. 2013, 2016a; and references therein).
The important results reported in these papers are: i) LBLs and IBLs show large amplitude IDV with high
duty cycles; ii) HBLs either don't show IDV or if show, the amplitude of variability and duty cycle of
IDV detection are less compare to LBLs and IBLs; iii) Cross-correlation analysis in different optical
bands on IDV timescales in general show strong correlation with zero lag which imply that the emission 
in different optical bands are co-spatial, but occasionally a few minutes time lags are also reported; 
iv) optical color variation on IDV timescales show a range of nature e.g. sometimes there is no color 
variation is seen, on some occasions BLLs show bluer when brighter (BWB) and FSRQs show redder when 
brighter trend (RWB), and rarely an opposite trend is also noticed; v) IR/optical SED in general fitted 
with a power-law but occasionally show big blue bump (BBB) which show the signature of thermal emission 
from accretion disk; vi) IDV LCs are used to get variability timescale which give clue of size of emitting 
region, black hole mass estimation of the blazar; etc. 

\subsection{IDV in Radio bands}

\noindent
IDV in radio bands are mixture of intrinsic and extrinsic origin. Mostly observations in centimeter and 
meter wavelengths have dominant extrinsic variability which is due to interstellar scintillation of radio
waves caused by the turbulent interstellar medium of the Milky Way while IDV in millimeter wavelength of 
intrinsic origin. Radio IDV of blazars are pioneered by 100m Effelsberg radio telescope in Germany, and
other radio and mm wavelengths telescopes (e.g. Wagner et al. 1990; Quirrenbach et al. 1991; Ostorero et 
al. 2006; Agudo et al. 2006; Gab$\acute{a}$nyi et al. 2007, 2009; Fuhrmann, et al. 2008; Marchili et al. 
2011, 2012; Gupta et al. 2012; Liu et al. 2013; Liu et al. 2012, 2015, 2017; and references therein). \\
\\
Wagner et al. (1990) did simultaneous optical and radio monitoring of blazars. Quirrenbach et al. (1991) 
reported for the first time the correlated optical and radio IDV in the blazar S5 0716+714. To test the
inverse-Compton (IC) catastrophe scenario in the blazar S5 0716+714 extensive observational campaign in radio 
and mm wavelengths were coordinated (Ostorero, et al. 2006; Agudo et al. 2006; Fuhrmann et al. 2008), and
the lower limits to brightness temperature was derived from the inter-day variations exceed the 10$^{12}$ K 
IC-limit by up to 2--4 orders of magnitude. Gab$\acute{a}$nyi et al. (2007) reported radio IDV of the blazar 
J 1128+5925 in three frequencies i.e. 2.7 and 10.45 GHz observations using 100m Effelsberg radio telescope 
in Germany, and 4.8 GHz observations
using 25m radio telescope in Urumqi, China. The observed frequency dependent IDV in the source was in good
agreement with prediction from interstellar scintillation. VLBA observation of the blazar J1128+592 is
reported by Gab$\acute{a}$nyi et al. (2009), and with VLBA observations they detected an east-west oriented 
core-jet structure with no significant motion in its jet. Radio IDV in blazars are studied where variability
characteristics have changed abruptly by interstellar scintillation (Marchili et al. 2011, 2013). Radio
IDV at 4.8 GHz using 25m Urumqi, China telescope for the blazars S5 0716+714 and 1156+295 are reported
by (Liu et al. 2012; Liu et al. 2013). Simultaneous IDV in X-ray, optical four bands and three frequencies
in radio are reported by Gupta et al. (2012). IDV detected in all three radio frequencies and also noticed
that low and high frequencies correlation does not peak at zero lag which show that low frequency radio
observation is combined effect of intrinsic and extrinsic mechanism. Optical and radio IDV observations
were carried out by Liu et al. (2017). IDV observation along with VLBI analysis is carried out for the 
blazar S4 0917+624 (Liu et al. 2015).    

\section{Quasi Periodic Oscillations in Intra-day Time Series Multi-wavelength Data}

\noindent
Detection of quasi-periodic oscillations (QPOs) in time series data are very rare in AGN. In last one 
decade, several detection of QPOs in AGN on diverse timescales ranging as short as few minutes and as 
long as few years using $\gamma-$ray, X-ray, optical and radio time series data are made (e.g. 
Gierli{\'n}ski et al.\ 2008; Espaillat et al.\ 2008; Gupta et al.\ 2009; Lachowicz et al.\ 2009; Rani 
et al.\ 2009, 2010; Lin et al.\ 2013; Sandrinelli et al.\ 2014, 2016a, 2016b, 2017; Graham et al.\ 2015; 
Ackermann et al.\ 2015; Alston et al.\ 2014, 2015; Pan et al.\ 2016; Bhatta et al.\ 2016b; Bhatta\ 2017; 
and references therein). But only a few claims of QPO detection on IDV timescales using X-ray and optical 
monitoring data of a few blazars are reported (Espaillat et al.\ 2008; Gupta et al.\ 2009; Lachowicz et 
al.\ 2009; Rani et al.\ 2010). \\
\\
There are several methods which can be used to search for QPO or periodic signal in time series data.
These methods are described by different groups and relevant source references are cited in their papers. 
Gierli{\'n}ski et al.\ (2008) used power spectral density (PSD) and data folding. Espaillat et al.\ (2008) 
used wavelet analysis, whereas Gupta et al.\ (2009) used wavelet plus randomization technique. Weighted 
wavelet z-transform (WWZ) and Lomb Scargle periodogram (LSP) are used by Bhatta (2017). Multiple analysis 
techniques e.g. structure function (SF), wavelet analysis, data folding, PSD, multi-harmonic AoV periodogram 
(mhAoV) are used by (Lachowicz et al.\ 2009). SF and auto correlation function (ACF) are used to get 
variability timescale and QPOs (Gaur et al. 2010; Pandey et al. 2017). One or multiple methods used in these 
papers are used in other searches for QPOs. \\
\\
QPOs in blazars on IDV timescales can be explained by several standard models of AGN. One of the simplest  
models in which the central BHs of AGN would attribute the QPOs can be explained by presence of a single 
dominating hot-spot on the accretion disk (e.g., Mangalam \& Wiita 1993; Chakrabarti \& Wiita 1993). Using 
QPO or nearly periodic signal, the period can be used to estimate the BH mass for non-rotating (Schwarzschild) 
BH, and maximally rotating (Kerr) BH. The detailed explanation is given in Gupta et al. (2009). Other 
alternative possible mechanisms for QPOs in blazars on IDV timescales can also have a disk origin or can arise 
from relativistic jets. The former class includes small epicyclic deviations in both radial and vertical  
directions from exact planar motions within a thin accretion disk (e.g., Abramowicz 2005), and trapped 
pulsational modes within a disk (e.g., Perez et al. 1997; Espaillat et al. 2008). Using detailed explanation
of Perez et al. (1997), one can also get the BH mass of the blazar. There are various jet models which also
can explain the QPO detection in blazars on IDV timescales e.g. a shock propagating down a jet in which jet 
structure is quasi-helical and change in electron density or magnetic field can produce QPO, a short lived
QPO can be due to turbulence behind the shock in the relativistic jet (e.g. Camenzind \& Krockenberger 1992; 
Gopal-Krishna \& Wiita 1992; Marscher et al. 1992).    
   
\subsection{In Optical} 

\noindent
Gupta et al.\ (2009) selected 20 optical IDV light curves of the blazar S5 0716+714 from a database of 102
lights curves taken in three years span (Montagni et al. 2006). They used wavelet analysis along with 
randomization test and found strong evidence for nearly periodic variations on 5 light curves with probability
$>$ 99\%. The period for these five light curves are found in the range of 25 minutes to 73 minutes which 
lead to BH mass ranging 2.47--7.35 $\times$ 10$^{6}$ M$_{\odot}$ and 1.57--4.67 $\times$ 10$^{7}$ M$_{\odot}$ 
for non-rotating BH and maximally rotating BH, respectively. Another evidence of QPO detection in optical band 
on the same blazar S5 0716+714 is reported by Rani et al. (2010). They found QPO period of $\sim$ 15 minutes
using various techniques (e.g. SF, LSP, PSD, data folding). This period yield the BH mass 
1.5 $\times$ 10$^{6}$ M$_{\odot}$ and 9.6 $\times$ 10$^{6}$ M$_{\odot}$ for non-rotating BH and maximally 
rotating BH, respectively. 
   
\subsection{In X-rays} 

\noindent
Using wavelet technique, (Espaillat et al.\ 2008) analyzed 19 observations of 10 AGN observed with EPIC/pn 
detector on board to XMM-Newton, and detected QPO period of 3.3 ks in one light curve of the blazar 3C 273. 
The QPO period is used to get the black hole (BH) mass of the blazar. They estimated the BH mass of the 
blazar is 7.3 $\times$ 10$^{6}$ M$_{\odot}$ and 8.1 $\times$ 10$^{7}$ M$_{\odot}$ for non-rotating BH and 
maximally rotating BH, respectively. In another observation of EPIC/pn of XMM-Newton for the blazar 
PKS 2155--304, (Lachowicz et al. 2009) detected QPO period of 4.6 h in which period was present for 
$\sim$ 3.8 cycles. This QPO detection was verified by various techniques (e.g. SF, PSD, MHAoV, data folding 
and wavelet). The BH mass of the blazar is estimated to be 3.29 $\times$ 10$^{7}$ M$_{\odot}$ and 
2.09 $\times$ 10$^{8}$ M$_{\odot}$ for non-rotating BH and maximally rotating BH, respectively. 

\section{Conclusions}

\noindent
With the extensive studies of IDV in radio to optical bands in last about three decades and in high 
energies (X-ray and $\gamma-$rays) in last one decade, we reach on the following conclusion: \\ 
\\
$\bullet$ Blazars show large amplitude IDV in radio bands which is basically the mixture of
extrinsic and intrinsic nature. \\
$\bullet$ LBLs and IBLs show large amplitude IDV in optical/IR bands with high duty cycle. \\
$\bullet$ HBLs either don't show optical/IR IDV or if show the amplitude is low and the duty
cycle is very less compare to LBLs/IBLs.   \\
$\bullet$ In general blazars don't show color variation on IDV timescales. But occasionally
it is seen. \\
$\bullet$ Optical inter-band cross-correlation show that in general there is no time lag in different
optical bands. On some occasions the time lag of a few minutes are reported in different optical bands. \\
$\bullet$ Optical/IR SEDs are well fitted with single power law. \\
$\bullet$ Sometimes optical/IR SEDs show big blue bump which is a signature of emission from accretion
disk.  \\
$\bullet$ In X-rays, HBLs show large amplitude IDV, and the duty cycle is high. \\
$\bullet$ In X-rays, LBLs/IBLs either don't show IDV or if show the amplitude is less. The duty cycle
of X-ray IDV for LBLs/IBLs are much less compare to duty cycles for HBLs. \\
$\bullet$ In general hard and soft X-ray LCs of blazars show strong cross-correlation with zero lag which
imply the emission in hard and soft bands are co-spatial. \\
$\bullet$ IDV timescales in different EM bands are used to get black hole mass of the blazars, size of
emitting region, and the Doppler factor $\delta$. \\
$\bullet$ In general sensitivity of very high energy $\gamma-$ray facilities are poor but occasionally
blazars are observed on time resolutions of a few minutes to few hours. The best time resolution $\gamma-$ray
light curves give high Doppler factor $\sim$ 100 for the blazar PKS 2155--304.  \\
$\bullet$ QPOs in blazars on IDV timescales are rare. \\
$\bullet$ Occasionally QPOs on IDV timescales are detected in a few blazars in X-ray and optical bands.

%%%%%%%%%%%%%%%%%%%%%%%%%%%%%%%%%%%%%%%%%%
\vspace{6pt} 

%%%%%%%%%%%%%%%%%%%%%%%%%%%%%%%%%%%%%%%%%%
%% optional

%%%%%%%%%%%%%%%%%%%%%%%%%%%%%%%%%%%%%%%%%%
%\acknowledgments{The work is partially supported by Indo-Poland ....}
%
%%%%%%%%%%%%%%%%%%%%%%%%%%%%%%%%%%%%%%%%%%
\authorcontributions{The author was asked by the journal to write the review article. The author planned
and wrote the article.} 

%%%%%%%%%%%%%%%%%%%%%%%%%%%%%%%%%%%%%%%%%%
\conflictofinterests{The author declares no conflict of interest.} 

%%%%%%%%%%%%%%%%%%%%%%%%%%%%%%%%%%%%%%%%%%
% Citations and References in Supplementary files are permitted provided that they also appear in the reference list here.
\bibliographystyle{mdpi}

\renewcommand\bibname{References}

%%%%%%%%%%%%%%%%%%%%%%%%%%%%%%%%%%%%%%%%%%
\end{document}